# Untrained, Physics-Informed Neural Networks for Structured Illumination Microscopy

Zachary Burns[1], Zhaowei Liu[1,2,3]*

[1] *Department of Electrical and Computer Engineering, University of California, San Diego,*
*9500 Gilman Drive, La Jolla, California 92093, USA*
[2] *Material Science and Engineering Program, University of California, San Diego,*
*9500 Gilman Drive, La Jolla, California 92093, USA*
[3] *Center for Memory and Recording Research, University of California, San Diego,*
*9500 Gilman Drive, La Jolla, California 92093, USA*

*Corresponding authors' email addresses: zhaowei@ucsd.edu

## Abstract:

In recent years there has been great interest in using deep neural networks (DNN) for super-resolution image reconstruction including for structured illumination microscopy (SIM). While these methods have shown very promising results, they all rely on data-driven, supervised training strategies that need a large number of ground truth images, which is experimentally difficult to realize. Additionally, such trained methods are prone to overfitting and often fail to generalize well to data that is out of the training distribution making them unsuitable as tools of discovery. For SIM imaging, there exists a need for a flexible, general, and open-source reconstruction method that can be readily adapted to different forms of structured illumination. We demonstrate that we can combine a deep neural network with the forward model of the structured illumination process to reconstruct sub-diffraction images without training data. The resulting physics-informed neural network (PINN) can be optimized on a single set of diffraction limited sub-images and thus doesn't require any training set. We show with simulated and experimental data that this PINN can be applied to a wide variety of SIM methods by simply changing the known illumination patterns used in the loss function and can achieve resolution improvements that match well with theoretical expectations.

# Introduction

In the past two decades, structured illumination microscopy (SIM) has become an increasingly popular super-resolution imaging method due to its relatively low illumination intensity levels and fast wide field imaging speeds[1-4]. SIM was first described by Gustafsson in his seminal paper and operates by using patterned illumination to shift high frequency spatial information into the low-frequency passband of a microscope[5]. By using a series of patterns, the ill-posed problem of super-resolution can be conditioned, and the high-resolution object recovered. The highest attainable spatial frequency with linear SIM reconstruction ($f_{SIM}$) can be described by:

$$f_{SIM} = f_{det} + f_{ill} \qquad (1)$$

where $f_{det}$ is the maximum spatial frequency of the detection optics and $f_{ill}$ is the maximum spatial frequency of the illumination patterns. Therefore, the main drawback of traditional linear SIM is that it only yields around a 2x resolution improvement when the illumination patterns used are diffraction limited. Traditional SIM therefore was unable to compete with other super-resolution methods such as STED[6] and STORM[7] in terms of resolution. If we assume that the numerical aperture of the detection optics is fixed, then the only way to increase SIM resolution is to increase the resolution of the illumination. New advances in the ability to generate sub-diffraction structured illumination have opened the door to further resolution improvement.

Sub-diffraction illumination patterns can be generated by either using a non-linear process or from near-field confinement via nanoscale structures. Non-linear SIM has been demonstrated using fluorophore saturation and showed a 3x or better resolution improvement[8-9]. Recently, it was demonstrated that a two-photon upconversion process can be used to generate similar non-linear patterns with much lower intensity[10]. In theory, if non-linear SIM includes enough higher-order harmonics and has sufficient signal to noise ratio it can achieve nearly unlimited spatial resolution, although a practical system only limits to the first few harmonics[8]. Another strategy to create high-resolution illumination patterns is to use near-field illumination which is not subject to the far-field diffraction limit and can have very large K-vectors. High refractive index waveguide-based SIM was demonstrated by using silicon nitride[11] and gallium phosphide[12]. The resolution improvement of this approach is however still limited by the availability of the high index materials. Advances in plasmonics, metamaterials and nanoscale fabrication have greatly increased the ability to design near-field light patterns in the past decade[13-16]. Plasmonic SIM, using either surface plasmon interference[17,18] or localized plasmonic resonator arrays[19-21], can boost the resolution improvement to around 3x. By combining traditional SIM and plasmonic SIM, such resolution improvement may be extended to 4x[22]. Recently hyperbolic metamaterials[23,24] and organic hyperbolic materials[25] have been used to create extremely high K-vector illumination patterns and have shown resolution improvements far beyond 4x. With these advances, linear SIM with resolution down to 30nm scales or better is now possible.

Traditionally, SIM images have been recovered using analytic reconstruction algorithms[5,8]. However, these algorithms include many hand-tuned parameters that affect the final reconstructed image, and it is often unclear for users how to choose them in order to achieve the image with the highest fidelity[26-28]. Additionally, such algorithms presently only exist for SIM and non-linear SIM with perfect periodic illumination patterns and not any other type of SIM illumination. In light of this, many near-field and non-periodic SIM methods have relied on the blind-SIM algorithm for image reconstruction which does not require knowledge of the illumination patterns[29,30]. Yet, this algorithm is fundamentally poorly conditioned and thus produces worse results than if the illumination patterns are known. Regarding ease of use for the microscopy community, only for linear SIM are there open-source code packages available[31,32]. Thus, it would be highly useful to the microscopy community if there were an open-source SIM reconstruction method that could be readily adapted to any type of SIM illumination with few parameters to tune.

Recently, a rapidly growing approach for SIM image reconstruction has been the use of deep neural networks (DNNs). DNNs have demonstrated impressive SIM reconstruction results and have been shown to reduce the number of needed sub-frames, image in extremely noisy conditions, improve axial resolution, and reduce artifacts[33-38].

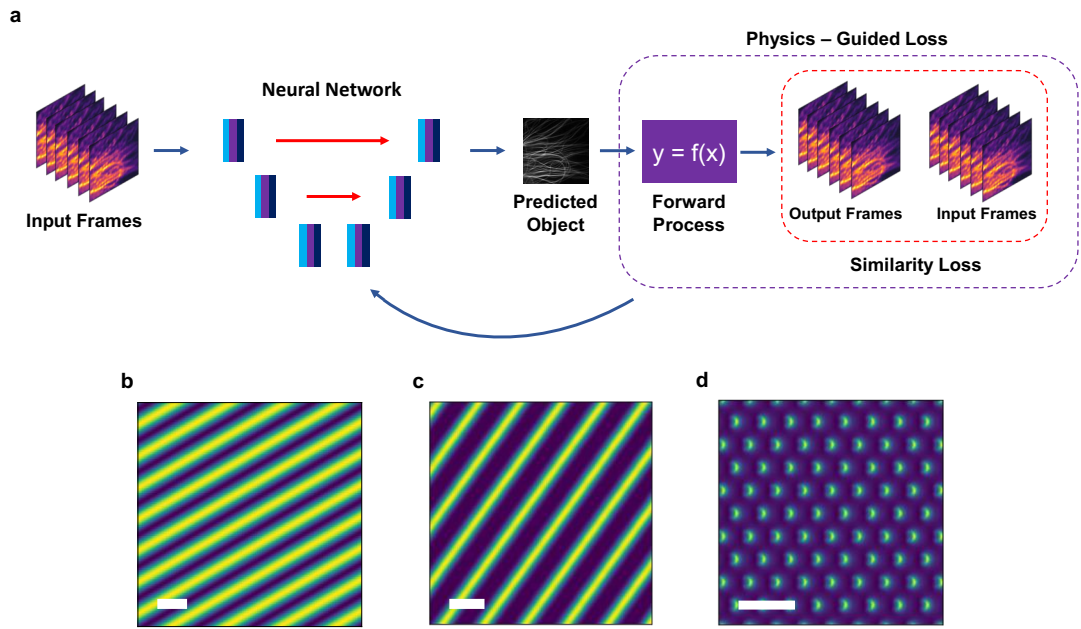

**Figure 1:** Concept of PINN for SIM. a) Flowchart for the PINN optimization process on a single set of sub-images. (b-d) Examples of SIM illumination patterns. a) traditional linear SIM b) non-linear SIM (NL-SIM) c) localized plasmonic SIM (LPSIM). White bars indicate the size of the diffraction limit relative to the patterns.

However, while DNNs have shown great promise, there are two main obstacles to their more practical and widespread use in the super-resolution community. The first is that all methods so far have employed a supervised learning strategy for training that requires paired low-resolution and high-resolution images. This is a challenge as is not easy to obtain experimental high-resolution ground truth images due to the diffraction limit. Thus, in order to get ground truth images another super-resolution technique is first needed, which as we explained earlier is not always available. The second major challenge is that data-driven supervised training methods can be highly sensitive to the class of objects used. These types of DNNs tend to have much worse performance when tried on objects that are different from the training set. Oftentimes they can hallucinate and project characteristic features of the training set onto the unseen objects[39,40]. It is highly impractical for a user to collect a new training set every time they want to try to use a DNN on a new type of object.

In this paper we propose a new reconstruction method that uses a DNN coupled with the forward model of the structured illumination process to produce super-resolution images without the need of a training set or ground-truth images. This so called "untrained" neural network is inspired by the deep image prior concept which has shown that DNNs can serve a type of general prior for natural images[41]. In recent years there has been growing interest in the use of untrained neural networks for various computational imaging problems[42-45]. However, to our knowledge, there has been no exploration of this concept as a method for the highly ill-posed problem of super-resolution microscopy where it could have a great advantage over traditional supervised DNNs. We demonstrate that this method is robust to a wide class of objects, can be used for many types of SIM illumination, can achieve resolution improvements close to the theoretical limit, and works well even in low signal to noise ratios.

## Theory

The physical forward model of incoherent structured illumination imaging in a microscope can be described as follows:

$$H(\rho) = (I\rho * PSF) + N \qquad (2)$$

Where $H$ is the diffraction limited sub-frame, $I$ is the illumination pattern, $\rho$ is the fluorophore distribution/object, $PSF$ is the point spread function of the microscope, and $N$ is additive noise. The goal of our reconstruction algorithm is to then find the object $f^*$ such that:

$$f^* = \arg\min_{f} \left\{ \sum_{i=1}^{n} \| H(\rho)_i - g_i \|_d + \alpha\varphi \right\} \qquad (3)$$

Where g is the experimentally collected sub-frame, $\| \|_d$ is a distance metric, $\alpha$ is a constant weighting term and $\varphi$ is a regularizer term or prior. This regularization term can either be

engineered by hand (e.g. sparsity penalty or total variation) or in the case of a data-driven DNN can be statistically learned during training. Note that this statistical prior is why data driven DNNs can have such impressive results but also why they tend to perform worse on data that varies from the training set.

For our untrained PINN we aim to instead learn the inverse mapping function M with trainable parameters (kernels) $\theta$ such that:

$$M_\theta = \arg\min_\theta \sum_{i=1}^{n} \left\| H(M_\theta(g_i)) - g_i \right\|_d \qquad (4)$$

The optimization process is shown visually in Figure 1. During the optimization process the neural network is fed the set of diffraction limited sub-images that are modulated via the SIM illumination patterns. The neural network then outputs an image which is then passed through the SIM forward process to generate a new series of diffraction limited sub-images. These images are then compared to the input frames and the loss backpropagated through the network to update the kernels. In this way the network is optimized without ever "seeing" a ground truth image. This process is repeated iteratively until the loss function plateaus and the network outputs the super-resolution image.

## Results

To test the performance of our PINN method we first conducted various tests using simulated data in order to have ground-truth images for proper evaluation. We assessed both the resolution improvement capabilities and versatility of the PINN across different SIM modalities and object types. We took experimental images of various biological objects from the BioSR dataset and used them as ground truth images[37,46] (Supp. 2). To simulate the diffraction limited subframes the images are multiplied by illumination patterns and then convolved with a known PSF which in our case is the Airy disc. The images are then down sampled to mimic pixilation and additive background noise is added to the images. For each case the SIM illumination patterns are used in the physics-informed loss function and training is run until the loss plateaus. We use 9 sub-frames (3 phases, 3 angles) for linear SIM, 25 sub-frames (5 phases, 5 angles) for NL-SIM, and 24 sub-frames (12 polar angles, 2 azimuthal angles) for LPSIM.

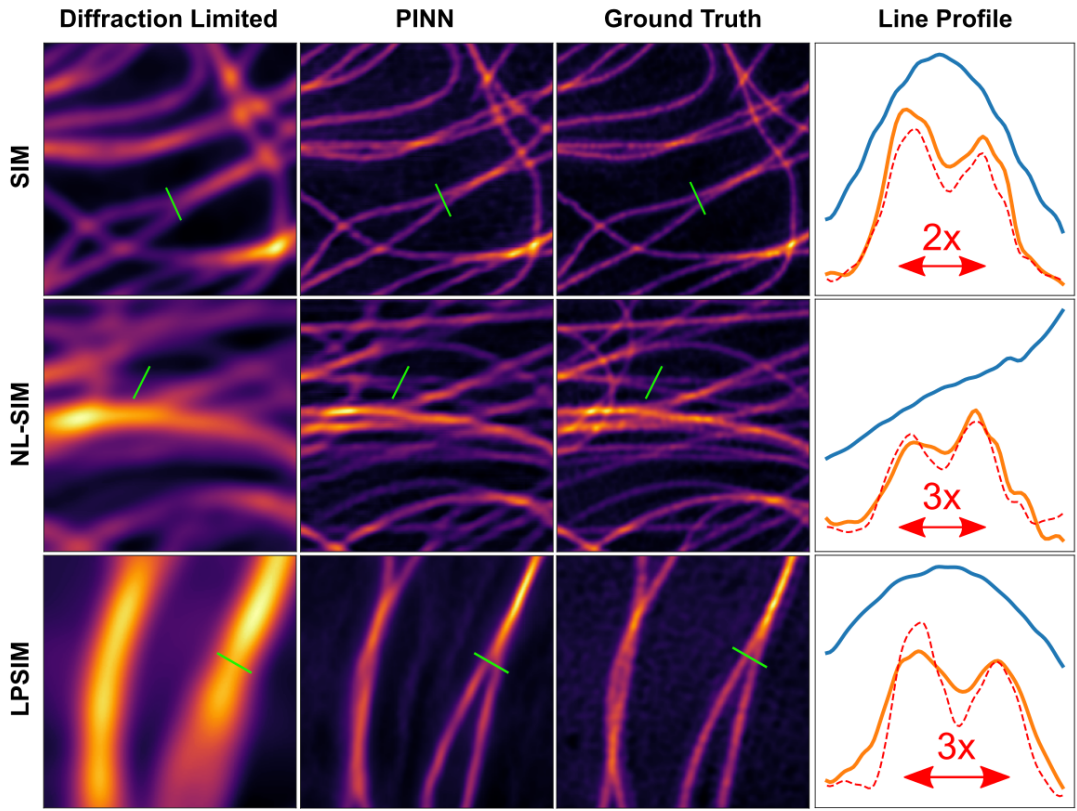

**Figure 2:** Demonstration of resolution improvement with multiple SIM modalities. Traditional linear SIM is able to distinguish features at 2x the diffraction limit and NL-SIM and LPSIM can accurately distinguish features at 3x the diffraction limit. Line Profiles: blue line = diffraction limited, orange line = PINN result, red dotted line = ground truth. SNR is 20 for all images.

We first asses the reconstruction ability of the PINN across a range of SIM illumination modalities. The results are shown in Fig. 2. For all three SIM imaging modalities (see Fig. 1b-d) there is clear resolution improvement and good agreement with the ground truth images. In terms of quality metrics (SSIM, PSNR, NRMSE) for all three modalities the values indicate close alignment between the PINN reconstructions and ground truth images (Supp. 3). To evaluate resolution, we compared line profiles across the images to determine at what distance sub-diffraction features can be resolved. In the linear SIM case, we are able to distinguish two microtubules at a distance 2x the diffraction limit. This confirms our expectation for linear SIM where the illumination patterns are also diffraction limited. Furthermore, with non-linear SIM and LPSIM we are able to resolve microtubules as distances 3x the diffraction limit, demonstrating that the PINN can be readily adapted to many types of SIM and that the resolution improvement of this method is fundamentally limited by the resolution of the illumination patterns used and how well-posed the reconstruction inverse process is. Additionally, the fact that the PINN resolution matches well with the resolution predicted for linear SIM by Equation 1 suggests that the network is truly learning the inverse problem rather than simply relying on object priors or performing deconvolution.

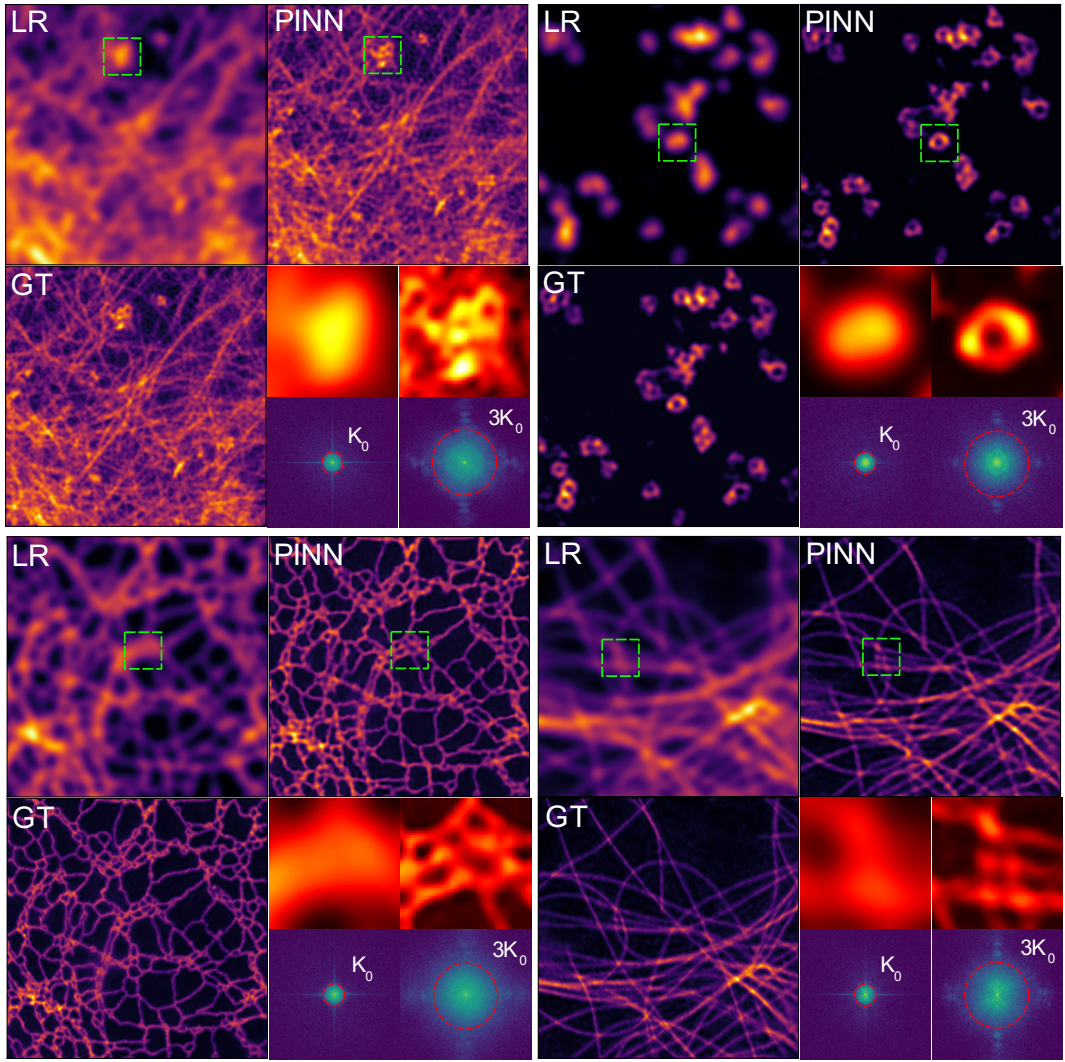

**Figure 3:** Demonstration of PINN based nonlinear SIM resolution improvement on multiple object types. (Top left) F-Actin, (top right) clathrin-coated pits, (bottom left) endoplasmic reticulum, (bottom right) microtubules. LR: low resolution (diffraction limited), PINN: physics-informed neural network, GT: ground truth. SNR is 20 for all images.

Next, we evaluated the ability of the PINN to reconstruct super-resolution images across a variety of biological objects. One of the drawbacks of trained neural networks is that they learn statistical priors about the objects in their training set which tends to limit their performance when tested on new types of objects. Super-resolution imaging is a tool of discovery, and thus it is important that a user has high confidence that the PINN can produce images on all types of images even if they have not been previously observed.

We test the PINN again on a series of simulated images that are generated from the BioSR dataset. The four object types tested are F-actin, clathrin-coated pits (CCP), endoplasmic reticulum, and microtubules, all of which have very different structures. If the PINN is simply biased to certain high-resolution structures (such as thin lines) rather than truly solving the inverse problem, then we would expect to observe hallucinations or artifacts. For our test we use non-linear SIM which presents a more challenging inverse problem as we aim to recover object features at more than 2x the diffraction limit.

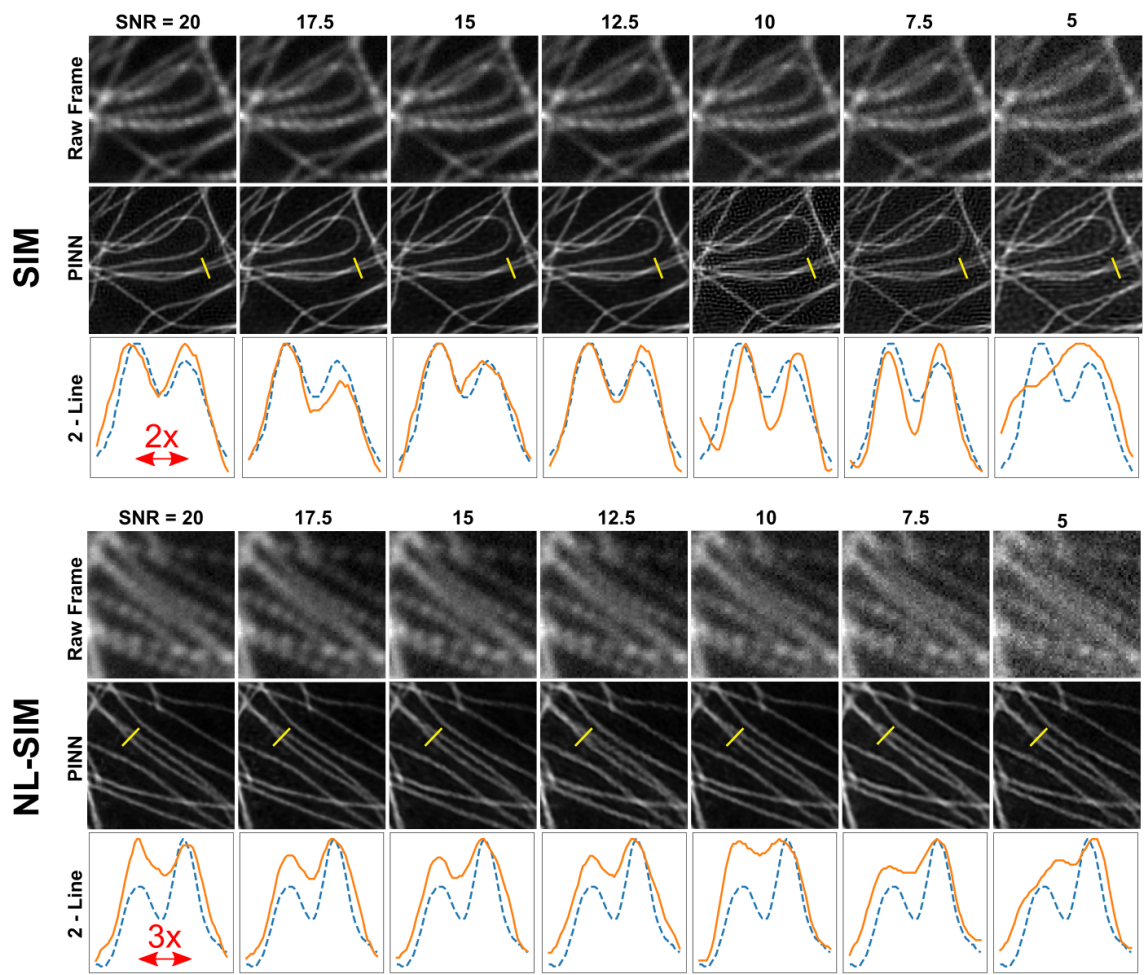

**Figure 4:** Assessment of PINN reconstruction performance at various signal to noise ratios. (Top section) SIM (bottom section) non-linear SIM. For each section: (top row) Individual raw frame at the given SNR, (middle row) The PINN reconstruction result, (bottom row) a line profile of two closely spaced microtubules (blue dotted) ground truth (orange) PINN result.

Results are shown in Figure 3. Across all objects we observe that the PINN produces images that match well with the ground truth images in terms of quality metrics (Supp 3). The insets clearly show sub-diffraction features and the PINN is able to recover them across a variety of objects. The frequency space images of the diffraction limited and PINN reconstructed are compared with circles indicating the range of support. The original circle indicates the diffraction limit in frequency space and the enlarged circle indicates a 3x enlargement. In all four object types the PINN result is able to increase the available frequency support to 3x the diffraction limit.

We further test the PINN by evaluating its performance at various signal to noise ratios (SNR). In practical microscopy the SNR will be limited by factors such as the exposure time and background noise, therefore it is important that our PINN method can work at reasonable SNRs. In figure 4 we evaluate the performance of the PINN on both linear and nonlinear SIM at SNRs ranging from 20 down to 5. We see that two-line features at 2x and 3x respectively are able to be distinguished down to a SNR of about 7.5 indicating that the PINN is able to recover super-resolution image features even at low SNR. Image quality metrics are included in Supplementary figure 4. For linear SIM the quality metrics remain high until the SNR of 12 and then drop off at 10. From Figure 4 it is evident that dot like noise artifacts begin to appear at a SNR of 10 corresponding to this drop in image quality. However, for nonlinear SIM the quality metrics remain stable with a very slight decrease down to an SNR of 5 indicating that it appears more resistant to noise related artifacts perhaps due to the larger number of input sub-frames used.

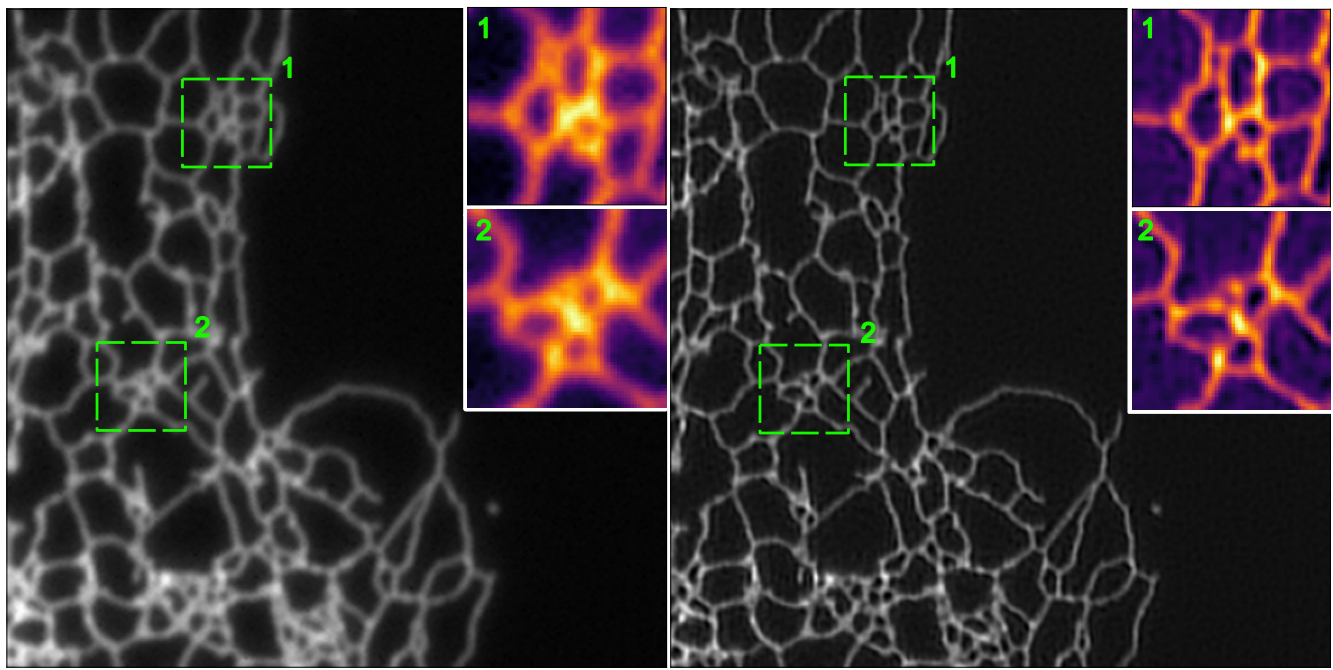

**Figure 5:** Experimental assessment of PINN for linear SIM on endoplasmic reticulum. (left) Diffraction limited image, (right) PINN result, (insets) Zoomed in view of dashed green regions showing sub-diffraction features.

Lastly, we test the PINN on experimental linear SIM data to confirm that the results match theory and our simulations. We use publicly available SIM data taken on samples of endoplasmic reticulum[38]. Since we do not have any true ground truth image in the case of experimental data, we used a decorrelation analysis tool to estimate resolution. The decorrelation analysis algorithm uses partial phase autocorrelation to measure resolution from a single image and has become a widely used tool for resolution estimation in microscopy[47]. Our results are shown in Figure 5. The PINN reconstruction shows clear resolution improvement as shown by both the zoomed in inset and the expansion of spatial frequency support. Additionally, the decorrelation analysis software package gives a resolution improvement of 1.78x for the PINN image which matches well with both the theoretical expected resolution improvement from Equation 1 and from the FairSIM ImageJ module (Supp. 6).

## Discussion and Conclusion

We demonstrate that an untrained physics-informed neural network can be used for the reconstruction of super-resolution images for structured illumination microscopy. The method does not require the collection of any training data and can be used for multiple types of SIM imaging modalities. The PINN is generalizable and shows high fidelity reconstruction across many types of biological objects and noise levels. Furthermore, the only hand-tuned parameter is the learning rate which typically produces good results when between $10^{-3}$ to $10^{-4}$.

As the PINN does not rely on a statistically learned prior, the main limiting factor of this method will be how ill-posed the inverse reconstruction problem is. If the SNR is too low, the modulation depth is poor, or the illumination patterns are not well chosen the reconstructed image will not be optimal.

Additionally, the current PINN method described assumes that both the illumination patterns and their location relative to the field of view are known. As demonstrated in Figure 5 this is experimentally possible, however, it can become challenging for cases where the illumination patterns are below the diffraction limit. Pattern location can be experimentally determined using markers like in the case of LPSIM. In future work, we aim to modify our method to account for the unknown pattern location in order to make reconstruction possible in cases where markers are difficult or not possible to fabricate.

We envision our PINN method as a flexible, open-source, and almost hyperparameter free method for general structured illumination imaging. We believe it will serve as an important tool for microscopists who need an easily modifiable reconstruction method that does not require any training data.

# Methods

**Simulated Image Generation**
Our simulated images are all created using data taken from the BioSR dataset (publicly available online)[46]. To generate the ground truth images crops are taken from high SNR SIM reconstructed results. The images are then convolved with a small PSF to smooth out any sharp artifacts. Widefield diffraction-limited sub-images are generated by multiplying the ground truth images with a series of illumination patterns. The product of the two are then convolved with the theoretical microscope PSF (airy disk). The sub frames are then down samples by 2x or 4x to mimic camera pixelization and additive Gaussian white noise is added. Illumination patterns for non-linear SIM are simulated assuming 3 harmonics[8] with 5 angles and 5 phases. Illumination patterns for LPSIM are based off full-wave simulations for nanopillar arrays[19-21] and have 12 polar angles and 2 azimuthal angles.

**Experimental Data**
The experimental SIM data was taken from publicly available data from the ML-SIM paper[38]. The data can be accessed at https://ml-sim.com/ and methods concerning experimental parameters are in the listed reference. Image resolution was performed using a ImageJ plugin of the decorrelation analysis algorithm[47] which is available at https://github.com/Ades91/ImDecorr.

**Neural Network Reconstruction**
The PINN is built using Tensorflow version 2.4.0 and Python version 3.8.5. The network uses a U-Net style architecture that has 3 downsampling and upsampling layers with skip connections between. We wrote a custom physics-informed loss function based on the SIM forward process and use the SSIM loss between the generated and input sub-frames to drive optimization. We use the Adam optimizer and a learning rate of between 0.001 and 0.0001. The learning rate is decayed exponentially with a decay rate of 0.9 every 50 epochs. In most cases we run the network optimization process for 1000 epochs which is more than enough for the loss to plateau. Running on a computer using a GTX 1080 Ti GPU and 64 Gb of RAM the reconstruction time can take from the order of minutes to tens of minutes depending on the size of the image, the size of the PSF kernel, and the number of sub-frames used.

**Acknowledge**
This work was supported by the National Science Foundation Graduate Research Fellowship Program (to Z. Burns) and by the Gordon and Betty Moore Foundation (to Z. Liu).

**Author Contributions**
Z.B. implemented software, ran experiments, and wrote the manuscript. Z.L. supervised the project and helped write the manuscript.

**Code and Data availability.** Code and data with demonstration Jupyter Notebooks for reconstruction are publicly available through GitHub: https://github.com/Zach-T-Burns/Untrained-PINN-for-SIM

**Competing interests:**

The authors declare no competing interests.

# References

1. Wu, Y., & Shroff, H. *Nature Methods* **15,** 1011–1019 (2018)
2. Saxena, M., Eluru, G., Gorthi, S. S. Structured illumination microscopy. *Advances in Optics and Photonics* **7**, 241-275 (2015).
3. Ströhl, F., & Kaminski, C. F., Frontiers in structured illumination microscopy. *Optica* **3**, 667-677 (2016).
4. Zheng, X., et al. Current challenges and solutions of super-resolution structured illumination microscopy. *APL Photonics* **6**, 020901 (2021).
5. Gustafsson, M. G. L. Surpassing the lateral resolution limit by a factor of two using structured illumination microscopy. *J. Microsc.* **198**, 82–87 (2000).
6. Hell, S. W. & Wichmann, J. Breaking the diffraction resolution limit by stimulated emission: stimulated-emission-depletion fluorescence microscopy. *Opt. Lett.* **19**, 780-782 (1994)
7. Rust, M. J., Bates, M. & Zhuang, X. Sub-diffraction-limit imaging by stochastic optical reconstruction microscopy (STORM). *Nature Methods* **3**, 793–796 (2006).
8. Gustafsson, M. G. L. Nonlinear structured-illumination microscopy: Wide-field flourescence microscopy with theoretically unlimited resolution. *Proc. Natl. Acad. Sci.* **102**, 13081-13086 (2005).
9. Rego, E. H., et al. Nonlinear structured-illumination microscopy with a photoswwitchable protein reveals cellular structures at 50-nm resolution. *Proc. Natl. Acad. Sci.* **109**, E135-E143 (2011).
10. Liu, B., et al. Upconversion Nonlinear Structured Illumination Microscopy. *Nano Letters* **20**, 4775-4781 (2020).
11. Tang, M., et al. High-Refractive-Index Chip with Periodically Fine-Tuning Gratings for Tunable Virtual-Wavevector Spatial Frequency Shift Universal Super-Resolution Imaging. *Adv. Sci.* **9**, 2103835 (2022).
12. Helle, Ø. I., et al. Structured illumination microscopy using a photonic chip. *Nature Photonics* 14, 431-438 (2020).
13. Ozbay, E. Plasmonics: Merging Photonics and Electronics at Nanoscale Dimensions. *Science* **311**, 189-193 (2006).
14. Gramotnev, D. K., Bozhevolnyi, S. I. Plasmonics beyond the diffraction limit. *Nature Photonics* **4**, 83-91 (2010).
15. Poddubny, A., Iorsh, I., Belov, P., & Kivshar, Y. Hyperbolic metamaterials. *Nature Photonics* **7**, 948-957 (2013).
16. Lorenzo, F., Wu, C., Lepage, D., Zhang, X., & Liu, Z. Hyperbolic metamaterials and their applications. *Prog. Quantum Electron.* **40**, 1-40 (2015).
17. Wei, F., & Liu, Z. Plasmonic structured illumination microscopy. *Nano Letters* **10**, 2531-2536 (2010).
18. Wei, F., et al. Wide Field Super-Resolution Surface Imaging through Plasmonic Structured Illumination Microscopy. *Nano Letters* **14**, 4634-4639 (2014).

19. Ponsetto, J., Wei, F., & Liu, Z. Localized plasmon assisted structured illumination microscopy for wide-field high-speed dispersion-independent super resolution imaging. *Nanoscale* **6**, 5807-5812 (2014).
20. Ponsetto, J., et al. Experimental demonstration of localized plasmonic structured illumination microscopy. *ACS Nano* **11**, 5344-5350 (2017).
21. Bezryadina, A., Zhao, J., Xia, Y., Zhang, X., & Liu, Z. High spatiotemporal resolution imaging with localized plasmonic structured illumination microscopy. *ACS Nano* **12**, 8248-8254 (2018).
22. Fernández-Domínguez, A. I., Liu, Z., & Pendry, J. B. Coherent four-fold super-resolution imaging with composite photonic-plasmonic structured illumination. *ACS Photonics* **2**, 341-348 (2015).
23. Lee, Y. U., et al. Metamaterial assisted illumination Nanoscopy via random super-resolution speckles. *Nat. Commun.* **12**, 1-8 (2021).
24. Lee, Y. U., et al. Ultrathin Layered Hyberbolic Metamaterial-Assisted Illumination Nanoscopy. *Nano Letters* **Article ASAP** (2022).
25. Lee, Y. U., et al. Organic Hyperbolic Material Assisted Illumination Nanoscopy. *Adv. Sci.* **8**, 2102230 (2021).
26. Karras, C., et al. Successful optimization of reconstruction parameters in structured illumination microscopy - A practical guide. *Opt. Comm.* **436**, 69-75 (2019).
27. Lal, A., Shan, C., & Xi, P. Structured Illumination Microscopy Image Reconstruction Algorithm. *IEEE. J. Sel. Top. In Quantum. Electron.* **22**, 50-63 (2016).
28. Smith, C. S., et al. Structured illumination microscopy with noise-controlled image reconstructions. *Nature Methods* **18**, 821-828 (2021).
29. Mudry, E., et al. Structured illumination microscopy using unknown speckle patterns. *Nature Photonics* **6**, 312-315 (2012).
30. Yeh, L. H., Tian, L., & Waller, L. Structured illumination microscopy with unknown patterns and a statistical prior. *Biomed. Opt. Express* **8**, 695-711 (2017).
31. Müller, M., Mönkemöller, V., Hennig, S., Hübner, W., & Huser, T. Open-source image reconstruction of super-resolution structured illumination microscopy data in ImageJ. *Nat. Commun.* **7**, 1-6 (2016).
32. Delp, S. L., et al. OpenSIM: Open-Source Software to Create and Analyze Dynamic Simulations of Movement. *IEEE Trans. Biomed. Eng.* **54**, 1940-1950 (2007).
33. Jin, L., et al. Deep learning enables structured illumination microscopy with low light levels and enhanced speed. *Nat. Commun.* **11**, 1-7 (2020).
34. Ling, C., et al. Fast structured illumination microscopy via deep learning. *Photonics Res.* **8**, 1350-1359 (2020).
35. Boland, M. A., Cohen, E. A. K., Flaxman, S. R., & Neil, M. A. A. Improving axial resolution in Structured Illumination Microscopy using deep learning. *Philos. Trans. Royal Soc. A.* **379**, 20200298 (2021).
36. Shah, Z. H., et al. Deep-learning based denoising and reconstruction of super-resolution structured illumination microscopy images. *Photonics Res.* **9**, B168-B181 (2021).

37. Qiao, C. Evaluation and development of deep neural networks for image super-resolution in optical microscopy. *Nature Methods* **18**, 194-202 (2021).
38. Chistensen, C. N., Ward, E. N., Lu, M., Lio, P., & Kaminski, C. F. ML-SIM: universal reconstruction of structured illumination microscopy images using transfer learning. *Biomed. Opt. Express* **12**, 2720-2733 (2021).
39. Hoffman, D. P., Slavitt, I. & Fitzpatrick, C. A. *Nature Methods* **18**, 131-132 (2021).
40. Belthangady, C., & Royer, L. A. *Nature Methods* **16**, 1215-1225 (2019).
41. Ulyanov, D., Vedaldi, A. & Lempitsky, V. Deep image prior. *Proc. IEEE Conf. Comput. Vis. Pattern Recognit.* 9446-9454 (2018).
42. Wang, F., et al. Phase imaging with an untrained neural network. *Light: Sci. Appl*. **9**, 1-7 (2020).
43. Bostan, E., Heckel, R., Chen, M., Kellman, M., & Waller, L. Deep phase decoder: self-calibrating phase microscopy with an untrained neural network. *Optica* **7**, 559-562 (2020).
44. Monakhova, K., Tran, V., Kuo, G., & Waller, L. Untrained networks for compressive lensless photography. *Opt. Express* **29**, 20913-20929 (2021).
45. Qiao, M. Liu, X., & Yuan X. Snapshot temporal compressive microscopy using an iterative algorithm with untrained neural networks. *Optics Letters* **8**, 1888-1891 (2021).
46. Qiao, C. & Li, D. BioSR: a biological image dataset for super-resolution microscopy.https://doi.org/10.6084/m9.figshare.13264793.v7(2020).
47. Descloux, A., Grußmayer, K. S., & Radenovic, A. Parameter-free image resolution estimation based on decorrelation analysis. *Nature Methods* **16**, 918-924 (2019).

# Supporting Information

# Untrained, Physics-Informed Neural Networks for Structured Illumination Microscopy

Zachary Burns[1], Zhaowei Liu[1,2,3]*

[1] *Department of Electrical and Computer Engineering, University of California, San Diego, 9500 Gilman Drive, La Jolla, California 92093, USA*
[2] *Material Science and Engineering Program, University of California, San Diego, 9500 Gilman Drive, La Jolla, California 92093, USA*
[3] *Center for Memory and Recording Research, University of California, San Diego, 9500 Gilman Drive, La Jolla, California 92093, USA*

*Corresponding authors' email addresses: zhaowei@ucsd.edu

## Contents

## S1. Neural network architecture

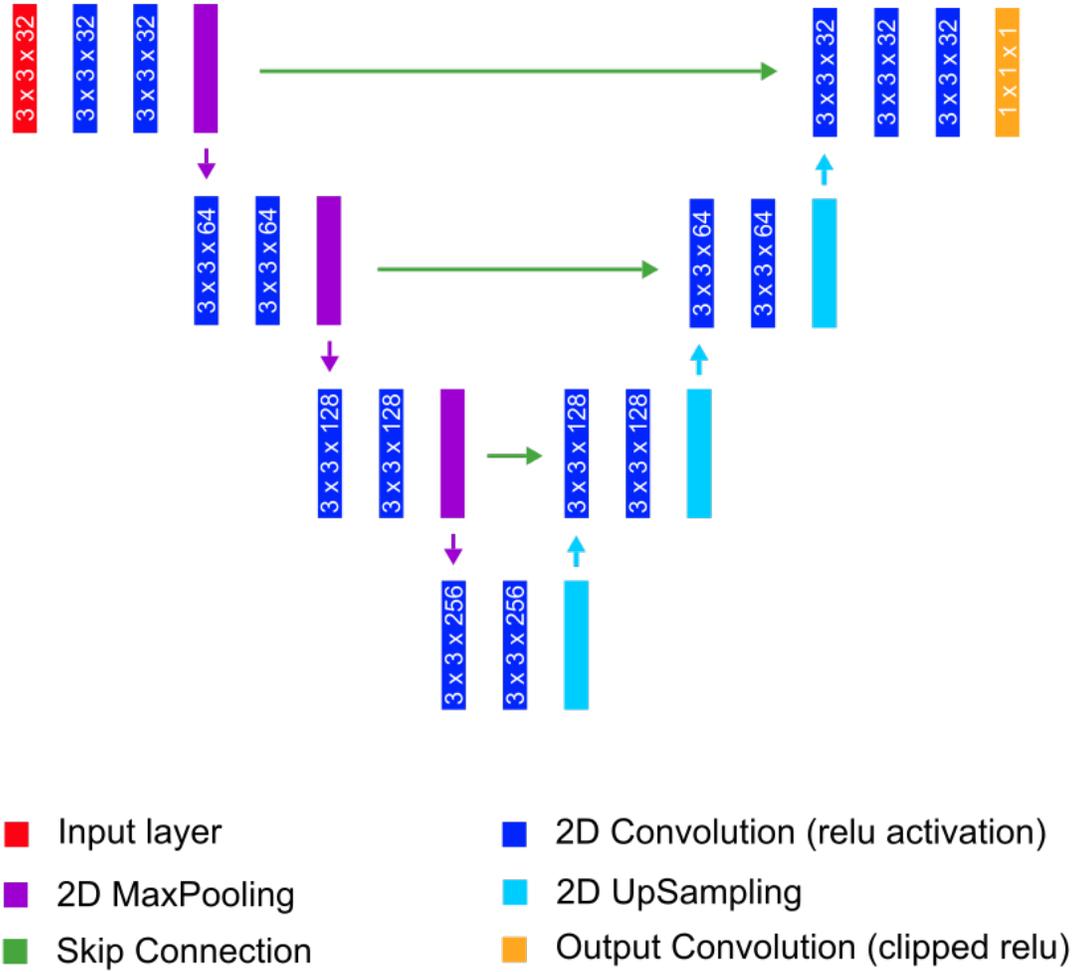

**Figure S1.** Diagram of the neural network architecture used in the paper. The network is based on the widely used "U-NET" structure which has an encoder/decoder architecture with residual skip connections. Numbers indicate the kernel size and number of kernels per convolutions layer. Example: 3x3x32 indicates 32 kernels of size 3x3.

## S2. Simulated data generation process

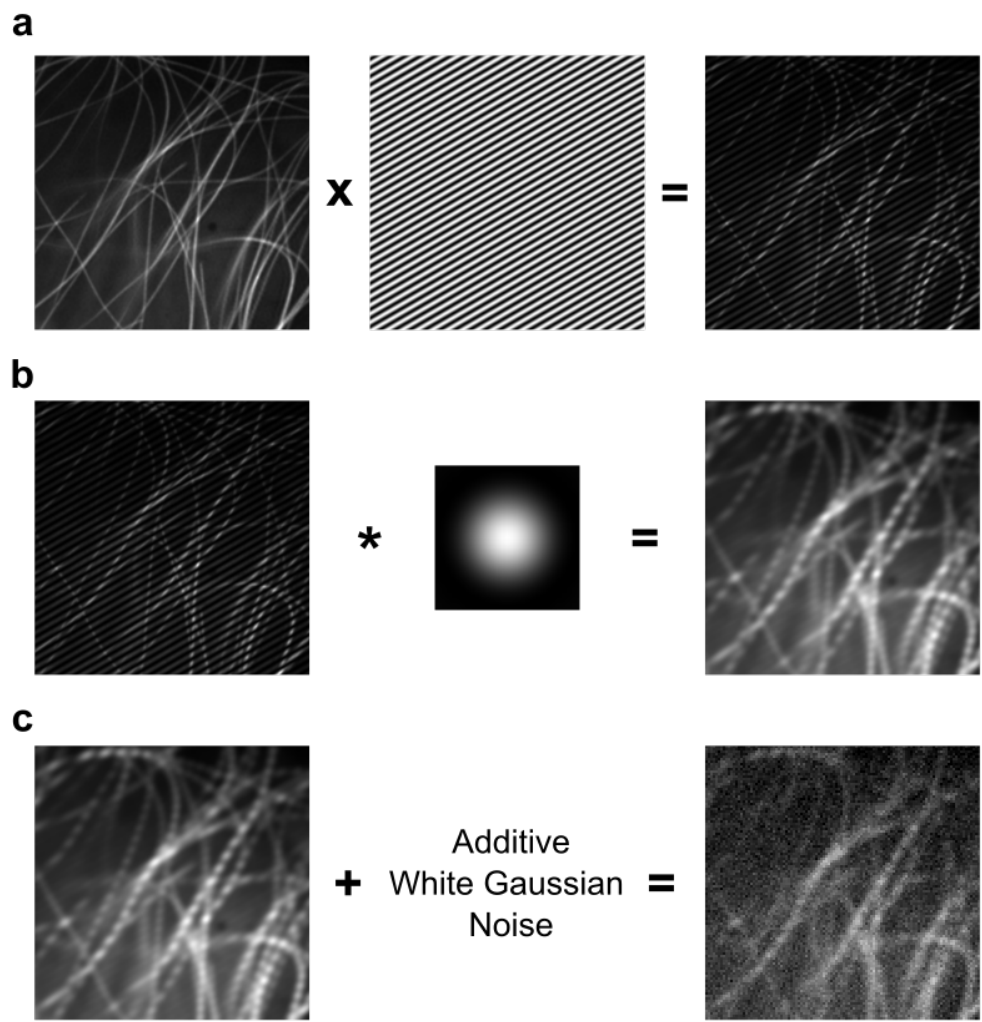

**Figure S2.** Process of generating simulated SIM images. (a) A "ground truth" image from the BioSR dataset is multiplied by the structured illumination pattern to get the effective illuminated object. (b) The effective illuminated area is convolved with the theoretical microscope PSF to get the diffraction limited sub-frame. (c) The image is down sampled several times to mimic the pixilation of the camera and additive white gaussian noise is added to mimic background noise.

## S3. Quantitative image metrics for Figure 2

|        | SIM    | NL-SIM | LPSIM  |
|--------|--------|--------|--------|
| PSNR   | 32.31  | 35.11  | 32.90  |
| SSIM   | 0.90   | 0.94   | 0.91   |
| NRMSE  | 0.0367 | 0.0263 | 0.0341 |

**Figure S3.** Quantitative image metrics for PINN reconstructed images in Figure 2 compared to the ground truth images used for data generation.

## S4. Quantitative image metrics for Figure 3

|        | F-Actin | CCP    | ER     | Microtubules |
|--------|---------|--------|--------|--------------|
| PSNR   | 29.42   | 34.06  | 25.82  | 29.64        |
| SSIM   | 0.81    | 0.85   | 0.86   | 0.87         |
| NRMSE  | 0.0878  | 0.1375 | 0.2234 | 0.1752       |

**Figure S4.** Quantitative image metrics for PINN reconstructed images in Figure 3 compared to the ground truth images used for data generation.

## S5. Quantitative image metrics for Figure 4

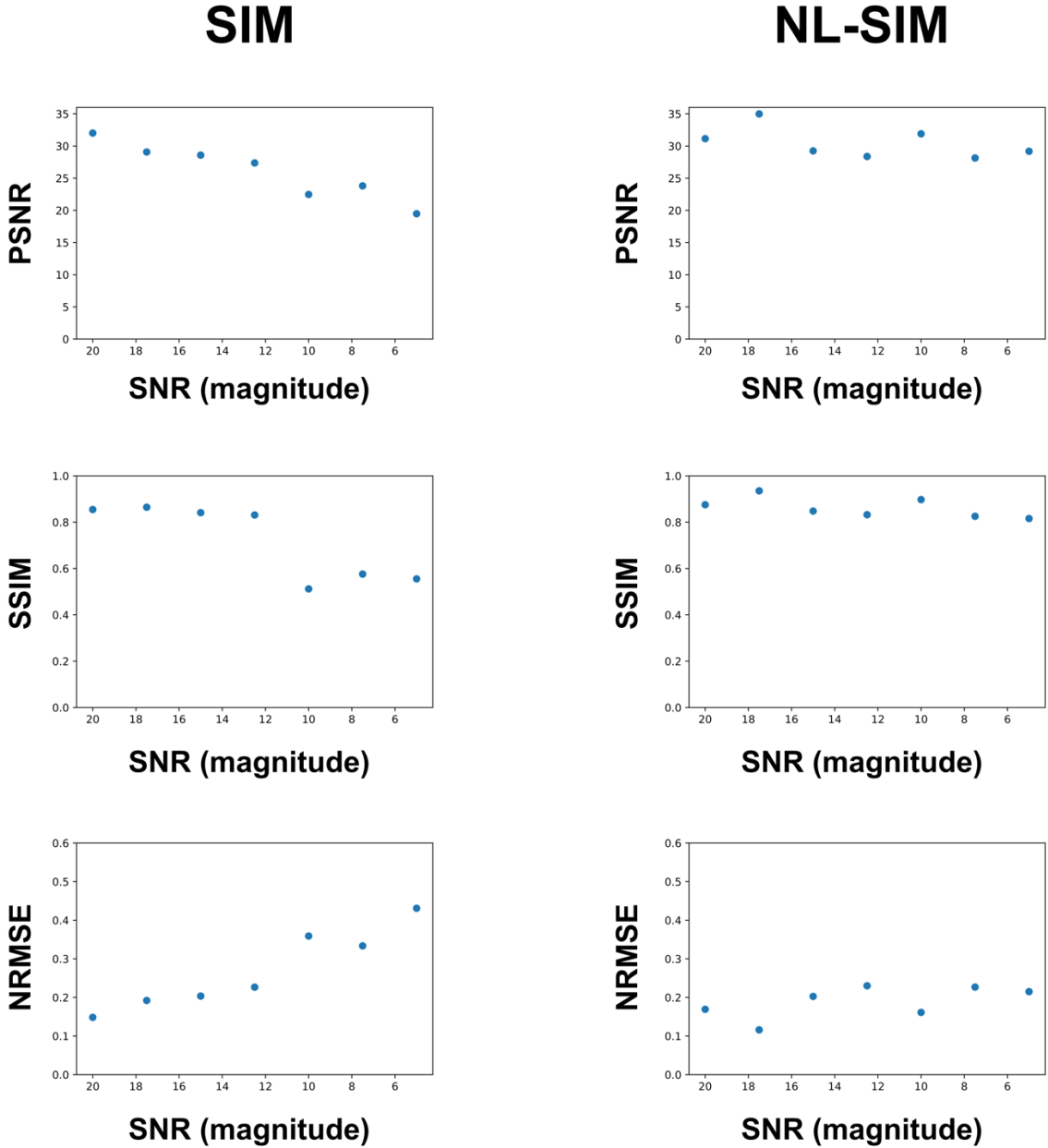

**Figure S5.** Quantitative image metrics (PSNR, SSIM, NRMSE) for PINN reconstructed images at various noise levels for linear SIM (left column) and nonlinear SIM (right column).

## S6. Additional information for experimental data

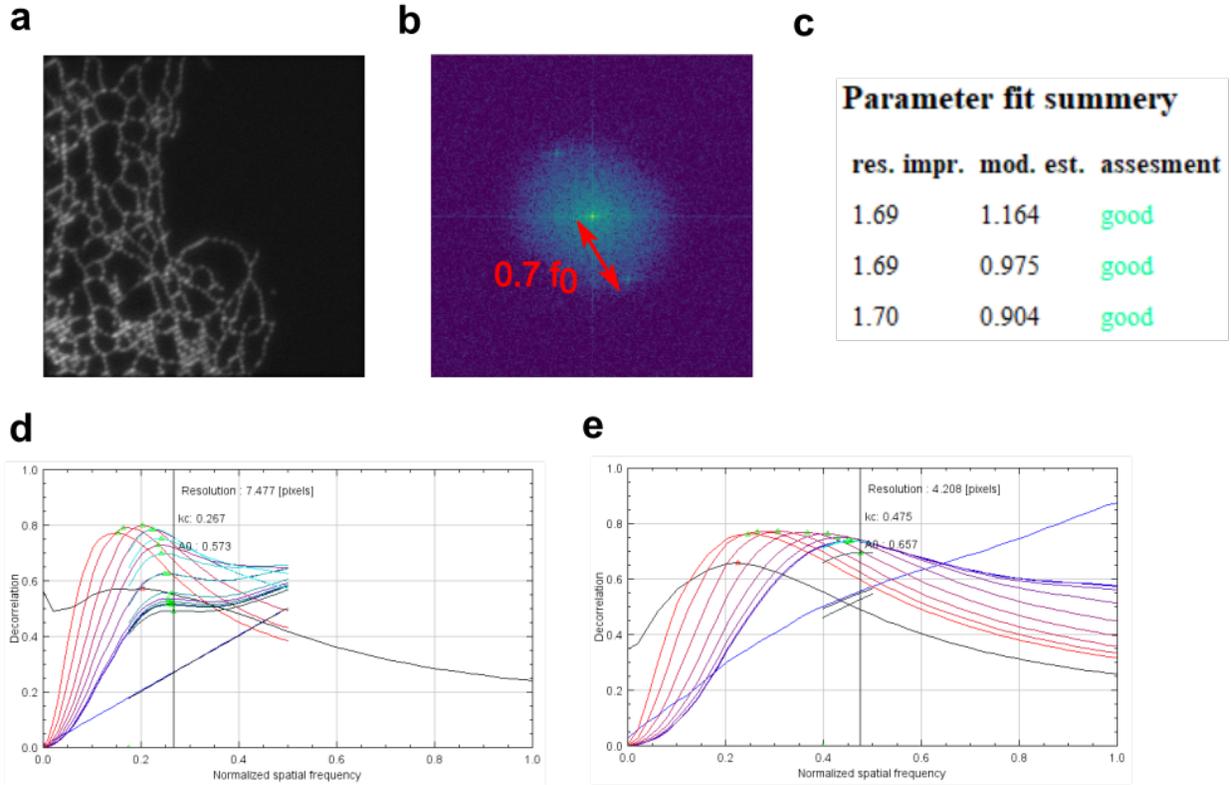

**Figure S6.** (a) One of the raw image frames from the experimental SIM data. (b) Frequency space view of the raw modulated frame. The twin intensity peaks correspond to the frequency of the illumination pattern and are approximately 0.7 times the diffraction limit spatial frequency. (c) Predicted resolution improvement from the FairSIM module. Image decorrelation analysis software results for (d) the diffraction limited image and (e) the PINN reconstructed image. The resolution of d divided by e gives about 1.7x which matches well with theory.